\documentclass[12pt]{iopart}
\usepackage{iopams,bm,cite,color,graphicx}

\expandafter\let\csname equation*\endcsname\relax
\expandafter\let\csname endequation*\endcsname\relax
\usepackage{amsmath}
\usepackage{youngtab}
\usepackage{caption}

\usepackage[symbol]{footmisc}

\newtheorem{example}{Example}[section]
\newtheorem{theorem}[example]{Theorem}
\newtheorem{corollary}[example]{Corollary}
\newtheorem{proposition}[example]{Proposition}
\newtheorem{conjecture}[example]{Conjecture}

\newcommand{\Omit}[1]{{\color{green}\textit{Omitted Stuff}}}
\newcommand{\Omitt}[1]{}

\def\ZZ{{\mathbb{Z}}}

\def\ZZpp{{\mathbb{Z}_{+}}}

\newcommand{\sgn}{\operatorname{sgn}}
\newcommand{\len}{\operatorname{len}}
\newcommand{\Det}{\operatorname{Det}}
\newcommand{\perm}{\operatorname{Perm}}
\newcommand{\id}{\operatorname{id}}
\newcommand{\imm}{\operatorname{Imm}}

\newcommand{\Proof}[1][]{\medskip\noindent {\it Proof#1: }}
\newcommand{\cqfd}{\hfill $\Box$\medskip}

\newcommand{\SymGrp}{\mathfrak S}
\newcommand{\DimGrp}[2]{\dim({#1},#2)}  
\newcommand{\HGrp}{\mathfrak H}
\newcommand{\VGrp}{\mathfrak V}

\newcommand{\norm}[1]{\lvert{#1}\rvert}

\newcommand{\calI}{{\mathcal I}}

\newcommand{\calA}{{\mathcal A}}
\newcommand{\calB}{{\mathcal B}}

\newcommand{\calY}{{\mathcal Y}}

\newcommand{\boldi}{\boldsymbol{i}}
\newcommand{\boldj}{\boldsymbol{j}}
\newcommand{\boldk}{\boldsymbol{k}}
\newcommand{\boldl}{\boldsymbol{\ell}}

\newcommand{\astrut}{\rule[-2.25mm]{0mm}{5.6mm}}
\newcommand{\dashline}{\leaders\hbox to 1.5pt{\hss
                          \vrule width0.75pt height0.4pt depth0.0pt}\hfill}

\newcommand{\A}[1]{{#1}^{+}}
\newcommand{\B}[1]{{#1}^{-}}

\usepackage{soul,xcolor}

\setstcolor{red}

\begin{document}

\title{Mixed symmetries of $\SymGrp_n$: 
             immanants in the sampling of $U(d)$ submatrices}

\author{Jacob~Daigle$^{1}$, Hubert~de~Guise$^{2}$ and Trevor Welsh$^{2}$\footnote{Authors listed in alphabetical order}}

\address{
$^1$Department of Mathematical Sciences, Lakehead University, Thunder Bay, Ontario, P7B 5E1, Canada,
$^2$Department of Physics, Lakehead University, Thunder Bay, Ontario, P7B 5E1, Canada.}

\eads{\mailto{tw3a23@soton.ac.uk}, \mailto{trevviewelsh@gmail.com}}

\begin{abstract}
We provide results on the mean and higher moments of immanants of submatrices
of ensembles of Haar-distributed unitary matrices, mainly without proofs.
This paper is based on a talk presented at ISQS29 in Prague in July 2025
by Trevor Welsh.
\end{abstract}



\vspace{-1cm}

\section{Introduction}

The quantum state of $n$ indistinguishable particles 
(bosons or fermions) must transform by a one-dimensional 
representation of $\SymGrp_n$,
the symmetric group on $n$ symbols
\cite{bach1988concept,kaplan2013pauli}.
This,
in turn, naturally introduces determinant and permanent states
for fermions and bosons respectively 
(see for instance
\cite{harvey1981symmetric,scheel2004permanents,mekonnen2025invariance}).
Applications of permanents in physics have been re-energized
by the \textsc{BosonSampling} problem \cite{aaronson2011computational},
which connects the computational complexity of sampling permanents of
$n\times n$ submatrices in the $d$-dimensional unitary group $U(d)$ ($n\ll d$)
to linear optics experiments.

For $n$ \emph{partially} distinguishable particles,
states with mixed permutation symmetries
(\emph{i.e.} states transforming under irreducible representations (irreps)
of $\SymGrp_n$ that are not fully symmetric or fully antisymmetric) may appear.
It is then natural to investigate \emph{immanants} either
in their own right \cite{merris1985inequalities,cash2003immanants},
for applications in chemistry \cite{cash2003immanants},
as a gateway to transition probabilities in $n$-body problems
\cite{dufour2024fourier}, or because they are sums of group functions
which appear in transition probabilities \cite{spivak2022generalized}.

In this contribution, we investigate moments of
immanants of $n\times n$ submatrices of random matrices in $U(d)$,
with particular emphasis on the first (the mean) and second moments.
They are of interest because they can be used
to understand the (anti)concentration property
of immanants of submatrices and thus provide
a pathway to prove average case complexity of these matrix
functions, thereby demonstrating a possible computational advantage
\cite{hangleiter2018anticoncentration,oszmaniec2022fermion,dalzell2022random}.  
Using Weingarten Calculus allows us to produce,
for the full range of immanants,
explicit $d$-dependent formulae for the mean,
and an algorithmic formula for the second moment.
We evaluate this latter formula for all immanants for $n\le 5$,
giving explicit rational polynomials in $d$,
with only the growth in computational requirements preventing us for going further.
Nonetheless, our analysis also yields a formula for the leading term,
which is evaluated here for $n\le 9$.
In the extremal cases of the immanant --- the permanent and determinant ---
our approach cogently reproduces results of Nezami \cite{Nezami2021}.
Full proofs of our results will be given elsewhere.


\section{Determinants, permanents and immanants}
\label{Sec:DPIs}

The \emph{determinant} $\Det M$ and \emph{permanent} $\perm M$
of an $n\times n$ matrix $M$ may be defined by
\begin{align}
\Det M&=\sum_{\pi\in \SymGrp_n}
\sgn(\pi)\,
M_{1\pi(1)} M_{2\pi(2)} \cdots M_{n\pi(n)} \label{Def:det}\\
\perm M&=\sum_{\pi\in \SymGrp_n}
M_{1\pi(1)} M_{2\pi(2)} \cdots M_{n\pi(n)} \label{Def:perm}
\end{align}
where $\sgn(\pi)=\pm1$ is the sign of the permutation $\pi\in\SymGrp_n$.

In the definition \eqref{Def:det} for the determinant,
the coefficient $\sgn(\pi)$ of each summand is the character
of the alternating representation of $\SymGrp_n$; 
in the definition \eqref{Def:perm}, the coefficients are all one
and so may be regarded as the values of the character
of the trivial representation of $\SymGrp_n$.

The \emph{immanants} interpolate this construction by taking the
coefficients to be values of the characters $\chi^\lambda$
of irreps of $\SymGrp_n$.
The irreps of $\SymGrp_n$ are labelled by partitions $\lambda$ of $n$.
Such partitions, denoted $\lambda\vdash n$,
are non-increasing sequences
$\lambda=(\lambda_1,\lambda_2,\ldots,\lambda_p)$ of positive
integer parts for which $\lambda_1+\lambda_2+\cdots+\lambda_p=n$.  
For later convenience, define $\len(\lambda)=p$.
Often, repeated parts of a partition are denoted using exponents:
for example, $(3,2^3,1^2)\equiv(3,2,2,2,1,1)$.
For $\lambda\vdash n$,
the immanant $\imm^{\lambda} M$ of an
$n\times n$ matrix $M$ is defined by
\begin{equation}\label{Def:Imm}
\imm^{\lambda} M=\sum_{\pi\in\SymGrp_n}
\chi^\lambda(\pi)
M_{1\pi(1)} M_{2\pi(2)} \cdots M_{n\pi(n)}\,.
\end{equation}
The partition $\lambda=(n)$ labels the trivial irrep of $\SymGrp_n$,
and so $\imm^{(n)}M=\perm M$.
The partition $\lambda=(1^n)$ labels the alternating irrep of $\SymGrp_n$,
and so $\imm^{(1^n)}M=\Det M$.

For $n=3$, the only partition other than $(1^3)$ or $(3)$ is $\lambda=(2,1)$, with corresponding immanant
\begin{equation}
\imm^{(2,1)}
\begin{pmatrix}
M_{11}&M_{12}&M_{13}\\
M_{21}&M_{22}&M_{23}\\
M_{31}&M_{32}&M_{33}
\end{pmatrix}
=
\begin{cases}
{(2)}\,M_{11}M_{22}M_{33}
+{(0)}\,M_{11}M_{23}M_{32}
+{(-1)}\,M_{12}M_{23}M_{31}\\
\quad
+{(0)}\,M_{13}M_{22}M_{31}
+{(-1)}\,M_{13}M_{21}M_{32}
+{(0)}\,M_{12}M_{21}M_{33}\,.
\end{cases}
\end{equation}
Here, for clarity, the values of the character $\chi^{(2,1)}$ have been
placed in parentheses.


\section{The mean}
\label{Sec:Mean}

In this work, for fixed $n\le d$, we consider $n\times n$ submatrices
$M$ of unitary matrices $U\in U(d)$.
We aim to understand the distribution of
$|{\imm}^\lambda M |^2$ as $U$ is sampled 
with respect to the Haar measure of $U(d)$.

The following result giving the mean of
$|{\imm}^\lambda M |^2$ is proved below in Section \ref{Sec:Wein}.
\begin{theorem}\label{Thrm:Mean}
For $\lambda\vdash n$ and $d\ge n$,
\begin{equation}\label{Eq:Mean}
\int_{U(d)} dU\,|{\imm}^\lambda M |^2
=
\frac{\DimGrp{\lambda}{\SymGrp_n}}{\DimGrp{\lambda}{U(d)}}\,.
\end{equation}
\end{theorem}
Eq.~\eqref{Eq:Mean} is the ratio of dimensions of irreps
of $\SymGrp_n$ and $U(d)$, both labelled by the \emph{same}
partition $\lambda$.
By using known expressions for these dimensions,
we can obtain a rational polynomial expression for
$\int_{U(d)} dU\,|\imm^\lambda M |^2$ for any $\lambda\vdash n$.

To give these dimension formulae, first define the \emph{Young diagram}
$\calY^\lambda$ of $\lambda\vdash n$ to be the following
subset of $\ZZ_{+}^2$:
\begin{equation}
\calY^\lambda=\{(i,j):1\le i\le\len(\lambda), 1\le j\le\lambda_i\}.
\end{equation}
The Young diagram $\calY^\lambda$ can be represented graphically
by placing a box at position $(i,j)$ for each $(i,j)\in\calY^\lambda$.
In the example $\lambda=(6,4,1)$, this gives:
\begin{center}
\yng(6,4,1)
\end{center}

For a partition $\lambda$, its conjugate partition $\lambda'$
is defined so that $\calY^{\lambda'}=\{(j,i):(i,j)\in\calY^{\lambda}\}$.
Thereupon, $\lambda'$ is obtained by reading the column lengths of
the diagram for $\calY^{\lambda}$.
For example, $(6,4,1)'=(3,2^3,1^2)$.

The \emph{hook-product} $H^\lambda$ of a partition $\lambda$ is
defined by
\begin{equation}
H^\lambda=\prod_{(i,j)\in\calY^\lambda}(\lambda_i+\lambda'_j-i-j+1).
\end{equation}
Often, the \emph{hook-length}
$\lambda_i+\lambda'_j-i-j+1$ is placed at position
$(i,j)$ for each box in $\calY^\lambda$.
In the following diagram, we have done this for $\lambda=(6,4,1)$.
$H^\lambda$ is simply the product of these values.
\begin{center}
\young(865421,5321,1)
\end{center}

The dimension of the representation of $\SymGrp_n$ labelled by $\lambda$ is then
\cite[\S2.37]{Robinson1961}
\begin{equation}\label{Eq:SymDim}
\DimGrp{\lambda}{\SymGrp_n}=\frac{n!}{H^\lambda}\,.
\end{equation}
For example, the dimension of the irrep of $\SymGrp_{11}$
labelled by $\lambda=(6,4,1)$ is
\begin{equation}
\DimGrp{(6,4,1)}{\SymGrp_{11}}
=\frac{11!}{8\times 6\times 5\times 4\times 2\times 5\times 3\times 2}
=\frac{11!}{57600}
=693\,.
\end{equation}

After defining the polynomial $N^\lambda(d)$ by
\begin{equation}\label{Eq:UniNum}
N^\lambda(d)=
\prod_{(i,j)\in\calY^\lambda} (d+j-i)
\end{equation}
the dimension of the representation of $U(d)$ labelled by
the partition $\lambda$ is given by
\cite[\S3.282]{Robinson1961}
\begin{equation}\label{Eq:UniDim}
\DimGrp{\lambda}{U(d)}=\frac{N^\lambda(d)}{H^\lambda}\,.
\end{equation}
Then, in our ongoing example,
\begin{equation}
N^{(6,4,1)}(d)
=d^2(d+1)^2(d+2)^2(d+3)(d+4)(d+5)(d-1)(d-2)
\end{equation}
so that the dimension of the irrep of $U(d)$
labelled by $\lambda=(6,4,1)$ is
\begin{align}
\DimGrp{(6,4,1)}{U(d)}
&=\frac{d^2(d+1)^2(d+2)^2(d+3)(d+4)(d+5)(d-1)(d-2)}{57600}\, .
\end{align}

Applying \eqref{Eq:SymDim} and \eqref{Eq:UniDim} to \eqref{Eq:Mean}
then yields
\begin{equation}\label{Eq:MeanB}
\int_{U(d)} dU\,|{\imm}^\lambda M|^2
=
\frac{n!}{N^\lambda(d)}\,.
\end{equation}
In particular, the $\lambda=(1^n)$ and $\lambda=(n)$ cases give%
\footnote{%
Nezami \cite{Nezami2021} also obtains these
(see his eqs.~(106) and (97)).}
\begin{gather}
\int_{U(d)} dU\,|\Det M |^2
=\frac{n!}{d(d-1)(d-2)\cdots(d-n+1)}
=
\binom{d}{n}^{-1}
\\
\int_{U(d)} dU\,|\perm M |^2
=\frac{n!}{d(d+1)(d+2)\cdots(d+n-1)}
=
\binom{d+n-1}{n}^{-1}.
\end{gather}
Some other cases are given in the second column of Table~\ref{Tab:Imm4}.

The following asymptotic result follows immediately from
\eqref{Eq:MeanB}:
\begin{corollary}\label{Cor:MeanAsmp}
For $\lambda\vdash n$,
\begin{equation}
\int_{U(d)} dU\,\vert {\imm}^\lambda M \vert ^{2}
=
\frac{n!}{d^{n}}
\,+\,O\left(\frac{1}{d^{n+1}}\right)\,.
\end{equation}
\end{corollary}


\section{Mean dominance}
\label{Sec:MeanDom}

\begin{figure}[h!]
\begin{minipage}[t]{.49\textwidth}
\centering
\caption{$\vert \textup{Imm}^{\lambda}\vert^2$ of $3\times 3$ submatrices
as a function of $d$: $(1^3)$ (circles), $(2,1)$ (triangles) and $(3)$ (diamonds).
Each data is averaged from $10^4$ submatrices of $d\times d$
Haar-random unitaries.
The thin lines are the predicted values where $d$ is taken as continuous.}
\label{Fig:partitions4}
\includegraphics[bb=0 0 450 277, scale=0.5]{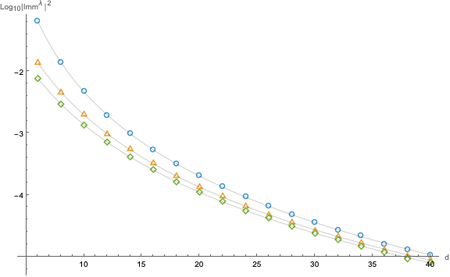}
\end{minipage}%
\hfill
\begin{minipage}[t]{.49\textwidth}
\centering
\caption{$\vert \textup{Per}\vert^4$ of $n\times n$ submatrices
as a function of $d$: $n=5$ (circles), $10$ (triangles) and $15$ (diamonds).  
Each data is averaged from $10^5$ submatrices of $d\times d$
Haar-random unitaries
The thin lines are the values predicted by Prop.\ref{Prop:Imm4}
with $d$ is taken as continuous.}
\label{Fig:conjecture}
\includegraphics[bb=0 0 450 277, scale=0.5]{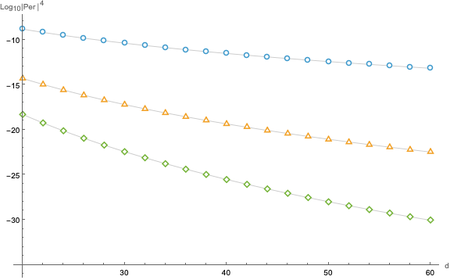}
\end{minipage}%
\end{figure}

Given partitions $\lambda,\mu\vdash n$, write $\lambda\trianglelefteq\mu$
if $\sum_{i=1}^k\lambda_i\le\sum_{i=1}^k\mu_i$
for $1\le k\le\min\{\len(\lambda),\len(\mu)\}$.
This defines a partial order on the set of partitions of $n$ known
as the \emph{dominance order}.
Note that, generally, it is not a total order, as illustrated
in \cite[\S1.4.7]{JamesKerber1981}.
Define $\lambda\lhd\mu$ if $\lambda\trianglelefteq\mu$
and $\lambda\ne\mu$.
It is easy to see \cite[\S1.4.9]{JamesKerber1981} that
if $\lambda\lhd\mu$,
then the diagram of $Y^\mu$ is obtained from that of $Y^\lambda$
by moving one or more boxes up and to the right.
Then, with $d$ fixed, \eqref{Eq:UniNum} gives $N^\lambda(d)<N^\mu(d)$.
Consequently, \eqref{Eq:MeanB} implies that:
\begin{proposition}\label{Prop:MeanDom}
If $\lambda,\mu\vdash n$ with $\lambda\lhd\mu$ then
for each $d\ge n$,
\begin{equation}\label{Eq:MeanDom}
\int_{U(d)} dU\,|\imm^\lambda M|^2
>
\int_{U(d)} dU\,|\imm^\mu M|^2\,.
\end{equation}
\end{proposition}
For the three partitions of $n=3$,
this result is illustrated in Fig.~\ref{Fig:partitions4}.


\section{Weingarten Calculus}
\label{Sec:Wein}

The proof of Theorem \ref{Thrm:Mean} makes use of the following result,
a cornerstone of \emph{Weingarten Calculus}
\cite{CMN2022}
\begin{theorem}\cite{Samuel1980,Collins2003}
\label{Thrm:Wein}
For fixed sequences
$\boldi=(i_1,\ldots,i_m)$,
$\boldj=(j_1,\ldots,j_m)$,
$\boldk=(k_1,\ldots,k_m)$,
$\boldl=(\ell_1,\ldots,\ell_m)$
of elements of the set $\{1,2,\ldots,d\}$,
\begin{equation}\label{Eq:Wein}
\int_{U(d)} dU\,U_{i_1k_1}\cdots U_{i_mk_m}
                U^\ast_{j_1\ell_1}\cdots U^\ast_{j_m\ell_m}
=
\sum_{\sigma,\tau\in\SymGrp_m}
\delta_{\sigma}(\boldi,\boldj)\,
\delta_{\tau}(\boldk,\boldl)\,
W(\sigma\tau^{-1},d)
\end{equation}
where
\begin{equation}
\delta_{\sigma}(\boldi,\boldj)=
\prod_{s=1}^m \delta_{i_{\sigma(s)},j_s}
\end{equation}
and for each $\rho\in\SymGrp_{m}$,
$W(\rho,d)$ is defined by
\begin{equation}\label{Def:WeinFun}
W(\rho,d)=
\frac{1}{(m!)^2}
\sum_{\xi\vdash m}
\frac{\DimGrp{\xi}{\SymGrp_{m}}^2}
     {\DimGrp{\xi}{U(d)}}\,
\chi^\xi(\rho)
\,.
\end{equation}
\end{theorem}

\Proof[ of Theorem \ref{Thrm:Mean}]
Using \eqref{Def:Imm} and the $m=n$ case of \eqref{Eq:Wein} yields:
\begin{equation}
\begin{split}
\int_{U(d)} dU\,|\imm^\lambda M|^2
&=\sum_{\pi,\gamma\in\SymGrp_n}
\chi^\lambda(\pi)
\chi^\lambda(\gamma)
U_{1\pi(1)}
\cdots U_{n\pi(n)}\,
U^\ast_{1\gamma(1)}
\cdots U^\ast_{n\gamma(n)}\,
\\
&=\sum_{\pi,\gamma\in\SymGrp_n}
\chi^\lambda(\pi)
\chi^\lambda(\gamma)
\sum_{\sigma,\tau\in\SymGrp_n}
\delta_{\sigma}(\boldi,\boldj)\,
\delta_{\tau}(\boldk,\boldl)\,
W(\sigma\tau^{-1},d)
\end{split}
\end{equation}
where 
$\boldi=\boldj=(1,2,\ldots,n)$,
$\boldk=\bigl(\pi(1),\pi(2),\ldots,\pi(n)\bigr)$
and
$\boldl=\bigl(\gamma(1),\gamma(2),\ldots,\gamma(n)\bigr)$.
Then, only $\sigma=\id$ and $\tau=\gamma\pi^{-1}$
contribute to the second sum above, resulting in
\begin{equation}
\begin{split}
\int_{U(d)} dU\,|\imm^\lambda M|^2
&=\sum_{\pi,\gamma\in\SymGrp_n}
\chi^\lambda(\pi)
\chi^\lambda(\gamma)
W(\pi\gamma^{-1},d)
\\
&=
\frac{1}{(n!)^2}
\sum_{\xi\vdash n}
\frac{\DimGrp{\xi}{\SymGrp_{n}}^2}
     {\DimGrp{\xi}{U(d)}}\,
\sum_{\pi,\gamma\in\SymGrp_n}
\chi^\lambda(\pi)
\chi^\lambda(\gamma)
\chi^\xi(\pi\gamma^{-1})
\,.
\end{split}
\end{equation}
The sum over $\gamma\in\SymGrp_n$ can be performed using an
extension of the first orthogonality property of finite group
characters (see \cite[eq.~(31.16)]{CurtisReiner1962}).
This gives
\begin{equation}
\begin{split}
\sum_{\pi,\gamma\in\SymGrp_n}
\chi^\lambda(\pi)
\chi^\lambda(\gamma)
\chi^\xi(\pi\gamma^{-1})
=
\delta_{\xi\lambda}\,
\frac{n!}{\DimGrp{\xi}{\SymGrp_{n}}}
\sum_{\pi\in\SymGrp_n}
\chi^\lambda(\pi)
\chi^\xi(\pi)
=
\delta_{\xi\lambda}\,
\frac{(n!)^2}{\DimGrp{\xi}{\SymGrp_{n}}}
\,.
\end{split}
\end{equation}
Applying this to the previous expression then yields
\eqref{Eq:Mean}, as desired.
\cqfd


\section{Higher Moments}

Moments, and in particular the second moment, are required in
applications to demonstrate approximate average-case hardness
of sampling through the anticoncentration of a output distribution
\cite{ehrenberg2025second}.
Thus, it is desirable to be able to calculate
\begin{equation}\label{Eq:Moments}
\int_{U(d)} dU\,|\imm^\lambda M |^{2t}
\end{equation}
for all $t\in\ZZpp$. Of course, \eqref{Eq:Mean} is the $t=1$ case.
However, evaluating \eqref{Eq:Moments} for $t>1$ is much trickier.
Here, we concentrate on the $t=2$ cases,
for which we have partial results.
To obtain these, we are led to consider Young subgroups of $\SymGrp_{2n}$.

\begin{figure}[h!]
\begin{minipage}[t]{.49\textwidth}
\centering
\caption{Demarking Young subgroups}
\label{Fig:YoungArray}
\centering
\includegraphics[scale=0.4]{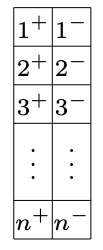}
\end{minipage}%
\hfill
\begin{minipage}[t]{.49\textwidth}
\centering
\caption{Action of $\epsilon_{\{2,3,5\}}\in\SymGrp_{12}$ \phantom{pad right}}
\label{Fig:YoungAct}
\includegraphics[scale=0.4]{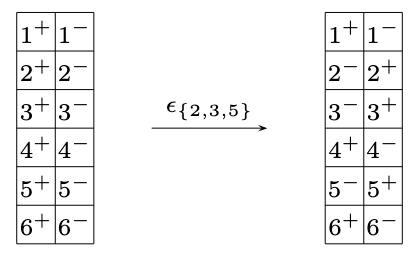}
\end{minipage}%
\end{figure}

First, let $\SymGrp_{2n}$ act to permute the set of symbols
$\calI_n=\{\A1,\A2,\ldots,\A{n},\B1,\B2,\ldots,\B{n}\}$,
and consider these symbols
arranged in a $n\times2$ array as in Fig.~\ref{Fig:YoungArray}.
Then, let $\VGrp_{n}$ denote the subgroup of $\SymGrp_{2n}$
which permutes the symbols within their columns in this array.
Each $\pi\in\VGrp_{n}$ is then composed from the two
permutations $\pi^+,\pi^-\in\SymGrp_n$ of the columns.
Denote this $\pi=\pi^+\oplus\pi^-$.
Note that $\norm{\VGrp_n}=(n!)^2$.

Similarly, let $\HGrp_{n}$ denote the subgroup of $\SymGrp_{2n}$
which keeps the symbols $\calI_n$ in their rows ---
it simply swaps the signs of some from the set $\{1,2,\ldots,n\}$.
Then, for each $\calA\subset\{1,2,\ldots,n\}$,
let $\epsilon_{\calA}\in\HGrp_{n}$ swap $\A{i}$ and $\B{i}$ for all $i\in\calA$.
For example, if $n=6$ and $\calA=\{2,3,5\}$, the action of
$\epsilon_{\calA}$ is depicted in Fig.~\ref{Fig:YoungAct}.
Note that $\norm{\HGrp_n}=2^n$.

Using the definition \eqref{Def:Imm} and then
applying the $m=2n$ case of the
Weingarten result \eqref{Eq:Wein} gives
\begin{equation}\label{Eq:Imm4byT}
\int_{U(d)} dU\,|\imm^\lambda M |^4
=\sum_{\xi\vdash2n}
\frac{1}{H^\xi\,N^\xi(d)}
\sum_{\begin{subarray}{c}
      \sigma,\rho\in\HGrp_n
     \end{subarray}}
T^{\lambda,\xi}(\sigma,\rho)
\end{equation}
with
\begin{equation}\label{Def:T4sum}
T^{\lambda,\xi}(\sigma,\rho)
=
\sum_{\begin{subarray}{c}
      \pi,\gamma\in\VGrp_n
     \end{subarray}}
\!
\hat\chi^\lambda(\pi)\,
\hat\chi^\lambda(\gamma)\,
\chi^\xi\bigl(\sigma\pi\rho\gamma)
\end{equation}
for $\sigma,\rho\in\HGrp_n$ and
$\hat\chi^\lambda(\A{\pi}\oplus\B{\pi})
   =\chi^\lambda(\A{\pi})\chi^\lambda(\B{\pi})$
for $\A{\pi},\B{\pi}\in\SymGrp_n$.

In the determinant case (where $\lambda=(1^n)$),
the double sum over $\HGrp_n$ and $\VGrp_n$
may be recognised as the (square of the) expression for
a \emph{Young symmetriser},
familiar from the representation theory of $\SymGrp_n$
\cite[\S1.5.10]{JamesKerber1981}.
This leads to:%
\footnote{In fact, the same reasoning gives
$\int_{U(d)} dU\,|\Det M |^{2t}
=
\bigl(\DimGrp{(t^n)}{U(d)}\bigr)^{-1}$.
Nezami \cite{Nezami2021} also gives this result
(see his eq.~(106)).}
\begin{proposition}\label{Prop:Det4}
\begin{equation}\label{Eq:Det4}
\int_{U(d)}\! dU\,|\Det M |^4
=
\frac{1}{\DimGrp{(2^n)}{U(d)}}\,.
\end{equation}
\end{proposition}


For the permanent case (where $\lambda=(n)$),
we have the following conjecture:%
\footnote{%
Nezami \cite{Nezami2021} makes a similar statement
(see his eq.~(98) which is derived from his conjectured eq.~(43)).}

\begin{conjecture}\label{Con:Perm4}
\begin{equation}\label{Eq:Perm4}
\int_{U(d)}\! dU\,|\perm M |^4
=
(n!)^2
\sum_{k=0}^{\lfloor\frac{n}{2}\rfloor}
2^{2n-4k}
\!\left(\!(2n-4k+1)\frac{(n-k)!}{k!(2n-2k+1)!}\right)^2
\!\!
\frac{1}{\DimGrp{(2n-2k,2k)}{U(d)}}
\,.
\end{equation}
\end{conjecture}

However, we have no such general expression for other $\lambda$,
and so we have resorted to evaluating our expression \eqref{Eq:Imm4byT}
for $\int_{U(d)} dU\,|\imm^\lambda M |^4$ using a computer.
Nonetheless, even for small $n$, this is temporally very expensive.
Indeed, to obtain the $n=5$ results, we have to be a bit smarter,
and use the fact that many of the
$T^{\lambda,\xi}(\sigma,\rho)$ are equal to one another.

First, for $\ell\ge0$, define the set
$[\ell]=\{1,2,\ldots,\ell\}$.
Then, for $\calA\subset[\ell]$, define
$[\ell]\backslash\calA=\{i\in[\ell],i\notin\calA\}$.
In addition, for later use, for $\ell\ge m\ge0$,
define $[\ell\backslash m]=[\ell]\backslash [m]\equiv\{m+1,m+2,\ldots,\ell\}$.


To express the symmetries of $T^{\lambda,\xi}(\sigma,\rho)$,
let $\calA,\calB\subset[n]$ be such that
$\sigma=\epsilon_{\calA}$ and $\rho=\epsilon_{\calB}$.
Then:
\begin{enumerate}
\item
$T^{\lambda,\xi}(\epsilon_{\alpha(\calA)},\epsilon_{\alpha(\calB)})=
T^{\lambda,\xi}(\epsilon_{\calA},\epsilon_{\calB})$
for all $\alpha\in\SymGrp_n$;
\item
$T^{\lambda,\xi}(\epsilon_{\calB},\epsilon_{\calA})=
T^{\lambda,\xi}(\epsilon_{\calA},\epsilon_{\calB})$;
\item
$T^{\lambda,\xi}(\epsilon_{[n]\backslash\calA},\epsilon_{[n]\backslash\calB})=
T^{\lambda,\xi}(\epsilon_{\calA},\epsilon_{\calB})$.
\end{enumerate}
Applying these three symmetries to the expression \eqref{Eq:Imm4byT}
leads to:%
\footnote{%
Rewriting our original naive expression in this way has made an
enormous efficiency improvement --- the original one
involved evaluating $2^{2n}$ terms
$T^{\lambda,\xi}(\sigma,\rho)$ for each $\xi$,
whereas \eqref{Eq:Imm4} makes approximately $(n+3)^3/24$ such evaluations.}
\begin{proposition}\label{Prop:Imm4}
For $\lambda\vdash n$,
\begin{equation}\label{Eq:Imm4}
\begin{split}
\int_{U(d)} dU\,|\imm^\lambda M |^4
&=\sum_{\xi\vdash2n}
\frac{1}{H^\xi\,N^\xi(d)}\\[-7pt]
&\qquad\times
\sum_{\ell=0}^n
\binom{n}{\ell}
\sum_{j=0}^\ell
\binom{\ell}{j}
\sum_{k=0}^{\min\left\{\substack{n-\ell-j\\
                            \ell-j}\right\}}
\!\!\!
\zeta_{n,\ell,k,j}
\binom{n-\ell}{k}
T^{\lambda,\xi}\left(\epsilon_{[\ell]},\epsilon_{[\ell+k\backslash \ell-j]}\right)
\end{split}
\end{equation}
where
$\zeta_{n,\ell,k,j}={4}/{(1+\delta_{k,n-\ell-j})(1+\delta_{k,\ell-j})}$.
\end{proposition}

The third column of
Table \ref{Tab:Imm4} gives evaluations of this result for
all $\lambda\vdash n$ with $n\le5$,
excluding those cases covered by Prop.~\ref{Prop:Det4}.

\begin{table}[h]
\caption{Evaluations of the 1st and 2nd moments}
\label{Tab:Imm4}
\footnotesize
\begin{center}
\begin{tabular}{|c|c|c|}
\hline
\astrut
$\lambda$&$\int_{U(d)} dU\,|\imm^\lambda M |^2$
         &$\int_{U(d)} dU\,|\imm^\lambda M |^4$\\
\hline
\astrut
(2)
&$\frac{2}{d(d+1)}$
&$4\,\frac{3d^2-d+2}{d^2(d^2-1)(d+2)(d+3)}$
\\
\multispan{3}\dashline
\\
\astrut
(3)
&$\frac{6}{d(d+1)(d+2)}$
&$144\,\frac{d^2+d+4}{d^2(d^2-1)(d+2)(d+3)(d+4)(d+5)}$\\
\astrut
(2,1)
&$\frac{6}{d(d^2-1)}$
&$36\,\frac{5d^3-3d^2-8d+12}{d^2(d^2-1)^2(d^2-4)(d+3)}$\\
\multispan{3}\dashline
\\
\astrut
(4)
&$\frac{24}{d(d+1)(d+2)(d+3)}$
&$576\,\frac{5d^4+30d^3+127d^2+294d+264}
              {d^2(d^2-1)(d+1)(d+2)^2(d+3)(d+4)(d+5)(d+6)(d+7)}$
\\
\astrut
(3,1)
&$\frac{24}{d(d^2-1)(d+2)}$
&$48\,\frac{73d^5+27d^4-585d^3-421d^2+742d-1240}
                {d^2(d^2-1)^2(d^2-4)(d+2)(d^2-9)(d+4)(d+5)}$
\\
\astrut
($2^2$)
&$\frac{24}{d^2(d^2-1)}$
&$144\,\frac{19d^4-112d^3+239d^2-224d+132}
                 {d^2(d^2-1)^2(d^2-4)^2(d^2-9)}$
\\
\astrut
(2,$1^2$)
&$\frac{24}{d(d^2-1)(d-2)}$
&$48\,\frac{73d^3-170d^2-151d+542}
                  {2d^2(d^2-1)^2(d-2)(d^2-4)(d^2-9)}$
\\
\multispan{3}\dashline
\\
\astrut
(5)
&$\frac{120}{d(d+1)(d+2)(d+3)(d+4)}$
&$28800\,\frac{3d^4+26d^3+173d^2+598d+880}
                 {d^2(d^2-1)(d+1)(d+2)^2(d+3)(d+4)(d+5)(d+6)(d+7)(d+8)(d+9)}$
\\
\astrut
(4,1)
&$\frac{120}{d(d^2-1)(d+2)(d+3)}$
&$960\,\frac{100d^6+581d^5+611d^4-3373d^3-9643d^2-7304d-12204}
                 {d^2(d^2-1)(d^2-4)(d^2-9)(d+1)(d+2)(d+3)(d+4)(d+5)(d+6)(d+7)}$
\\
\astrut
(3,2)
&$\frac{120}{d^2(d^2-1)(d+2)}$
&$240\,\frac{394d^6-367d^5-3331d^4+7568d^3-10747d^2+17575d+1428}
                   {d^3(d^2-1)^2(d^2-4)^2(d^2-9)(d+3)(d+4)(d+5)}$
\\
\astrut
(3,$1^2$)
&$\frac{120}{d(d^2-1)(d^2-4)}$
&$240\,\frac{418d^5-813d^4-5424d^3+7276d^2+18977d-35968}
                   {d^2(d^2-1)^2(d^2-4)^2(d^2-9)(d^2-16)(d+5)}$
\\
\astrut
($2^2$,1)
&$\frac{120}{d^2(d^2-1)(d-2)}$
&$240\,\frac{394d^5-3725d^4+12404d^3-15268d^2+609d+9540}
                   {d^3(d^2-1)^2(d^2-4)^2(d^2-9)(d-3)(d-4)}$
\\
\astrut
(2,$1^3$)
&$\frac{120}{d(d^2-1)(d-2)(d-3)}$
&$960\,\frac{100d^3-433d^2-47d+1638}
                     {d^2(d^2-1)^2(d^2-4)(d-2)(d^2-9)(d-3)(d-4)}$

\\
\hline
\end{tabular}
\end{center}
\end{table}


\section{Taking it to the limit}
\label{Sec:Limit4}

From \eqref{Eq:Imm4byT}, we obtain:
\begin{equation}\label{Def:Imm4lead}
\int_{U(d)} dU\,|\imm^\lambda M |^{4}
=
\frac{1}{d^{2n}}
{J^{\lambda}}
\,+\,
O\left(\frac{1}{d^{2n+1}}\right)
\end{equation}
where the leading term coefficient $J^\lambda\in\ZZ$ is given by
\begin{equation}\label{Def:Imm4leadbyJ}
{J^{\lambda}}
=
\sum_{\begin{subarray}{c}
      \sigma,\rho\in\HGrp_n
     \end{subarray}}
{J^{\lambda}(\sigma,\rho)}
\qquad
\text{with}
\qquad
{J^{\lambda}(\sigma,\rho)}
=
\sum_{\begin{subarray}{c}
      \pi,\gamma\in\VGrp_{n}\\
      {\sigma\pi\rho\gamma=\id}
      \end{subarray}}
\!\!
\hat\chi^\lambda(\pi)\,
\hat\chi^\lambda(\gamma)\,.
\end{equation}

Using this, or Prop.~\ref{Prop:Det4}, we obtain
\begin{equation}\label{Def:Det4leadbyJ}
{J^{(1^n)}}=n!(n+1)!
\end{equation}
Moreover, \eqref{Def:Imm4leadbyJ} may be used to show that
$J^{\lambda}=J^{\lambda'}$, where $\lambda'$ is the
conjugate of the partition $\lambda$.
Therefore, \eqref{Def:Det4leadbyJ} also gives
\begin{equation}\label{Def:Perm4leadbyJ}
{J^{(n)}}=n!(n+1)!
\end{equation}

For other cases of $\lambda$, we again resort to using a computer,
but not before we extract the symmetries
of $J^{\lambda}(\sigma,\rho)$.
To express these, again let $\calA,\calB\subset[n]$ be such that
$\sigma=\epsilon_{\calA}$ and $\rho=\epsilon_{\calB}$.
Then:
\begin{enumerate}
\item
$J^{\lambda}(\epsilon_{\alpha(\calA)},\epsilon_{\alpha(\calB)})=
J^{\lambda}(\epsilon_{\calA},\epsilon_{\calB})$
for all $\alpha\in\SymGrp_n$;
\item
$J^{\lambda}(\epsilon_{\calA},\epsilon_{\calB})=0$
if $\norm{\calA}\ne\norm{\calB}$;
\item
$J^{\lambda}(\epsilon_{[n]\backslash\calA},\epsilon_{[n]\backslash\calB})=
J^{\lambda}(\epsilon_{\calA},\epsilon_{\calB})$.
\end{enumerate}
The second symmetry of $T^{\lambda,\xi}(\sigma,\rho)$ also extends to
this case, but that given here is more powerful.

Applying the above three symmetries to the expression \eqref{Def:Imm4leadbyJ}
leads to the first part of the following:
\newcommand{\thetatwo}[2]{\theta_{#1,#2}}
\begin{proposition}\label{Prop:Imm4leadbyJ}
For $\lambda\vdash n$,
\begin{equation}\label{Eq:Imm4leadbyJ}
{J^{\lambda}}
=
\sum_{\ell=0}^{\lfloor\frac{n}{2}\rfloor}
\frac{2}{1+\delta_{2\ell,n}}
\binom{n}{\ell}
\sum_{k=0}^{\ell}
\binom{\ell}{k}
\binom{n-\ell}{k}
J^{\lambda}\bigl(\epsilon_{[\ell]},\epsilon_{[\ell+k\backslash k]}\bigr)\,.
\end{equation}
In addition,
\begin{equation}\label{Eq:Imm4leadbyJb}
J^{\lambda}\bigl(\epsilon_{[\ell]},\epsilon_{[\ell+k\backslash k]}\bigr)
=
\!\!
\!\!
\!\!
\sum_{\begin{subarray}{c}
      \alpha^+,\beta^+\in\SymGrp_{\ell}\\
      \alpha^-,\beta^-\in\SymGrp_{n-\ell}
      \end{subarray}}
\!\!
\!\!\!
\chi^\lambda\bigl(\thetatwo{\ell}{k}(\alpha^+\!\oplus\alpha^-)\bigr)\,
\chi^\lambda\bigl(\thetatwo{\ell}{k}(\beta^+\!\oplus\beta^-)\bigr)\,
\chi^\lambda\bigl(\thetatwo{\ell}{k}(\beta^+\!\oplus\alpha^-)\bigr)\,
\chi^\lambda\bigl(\thetatwo{\ell}{k}(\alpha^+\!\oplus\beta^-)\bigr)
\end{equation}
where $\thetatwo{\ell}{k}\in\SymGrp_n$ is the involution defined by
\begin{equation}
\thetatwo{\ell}{k}(i)=
\begin{cases}
i+\ell&\text{if $1\le i\le k$,}\\
i-\ell&\text{if $\ell+1\le i\le \ell+k$,}\\
i&\textup{otherwise.}
\end{cases}
\end{equation}
\end{proposition}

Using this result gives the values in Table~\ref{Tab:Imm4lead}.
Because $J^{\lambda}=J^{\lambda'}$, we only list one value
from each conjugate pair.

\begin{table}[h]
\caption{Evaluations of the leading term coefficient $J^\lambda$}
\label{Tab:Imm4lead}
\footnotesize
\begin{center}
\begin{tabular}{|c|r|}
\hline
\rule[-3.5pt]{0ex}{12pt}
$\lambda$&$J^\lambda$\\
\hline
(1)&2\\
\multispan{2}\dashline\\
(2)&12\\
\multispan{2}\dashline\\
(3)&144\\
(2,1)&180\\
\multispan{2}\dashline\\
(4)&2880\\
(3,1)&3504\\
(2,2)&2736\\
\multispan{2}\dashline\\
(5)&86400\\
(4,1)&96000\\
(3,2)&94560\\
(3,$1^2$)&100320\\
\hline
\end{tabular}
\qquad
\begin{tabular}{|c|r|}
\hline
\rule[-3.5pt]{0ex}{12pt}
$\lambda$&$\hfil J^\lambda\hfil$\\
\hline
(6)&3628800\\
(5,1)&3772800\\
(4,2)&4013280\\
(4,$1^2$)&3754080\\
($3^2$)&2895840\\
(3,2,1)&7128000\\
\multispan{2}\dashline\\
(7) & 203212800\\
(6,1) & 200793600\\
(5,2) & 205309440\\
(5,$1^2$) & 189987840\\
(4,3) & 184917600\\
(4,2,1) & 407090880\\
(4,$1^3$) & 185401440\\
($3^2$,1) & 190411200\\
\hline
\end{tabular}
\qquad
\begin{tabular}{|c|r|}
\hline
\rule[-3.5pt]{0ex}{12pt}
$\lambda$&$\hfil J^\lambda\hfil$\\
\hline
(8) & 14631321600\\
(7,1) & 13886208000\\
(6,2) & 13501071360\\
(6,$1^2$) & 12628869120\\
(5,3) & 13266247680\\
(5,2,1) & 25358135040\\
(5,$1^3$) & 11890851840\\
($4^2$) & 9760020480\\
(4,3,1) & 24651244800\\
(4,$2^2$) & 13852258560\\
(4,2,$1^2$) & 29371910400\\
($3^2$,2) & 12037536000\\
\hline
\end{tabular}
\qquad
\begin{tabular}{|c|r|}
\hline
\rule[-3.5pt]{0ex}{12pt}
$\lambda$&$\hfil J^\lambda\hfil$\\
\hline
(9) & 1316818944000\\
(8,1) & 1209522585600\\
(7,2) & 1123058442240\\
(7,$1^2$) & 1065369231360\\
(6,3) & 1107173007360\\
(6,2,1) & 1949153310720\\
(6,$1^3$) & 973714452480\\
(5,4) & 985895608320\\
(5,3,1) & 2449895777280\\
(5,$2^2$) & 1120506670080\\
(5,2,$1^2$) & 2257223915520\\
(5,$1^4$) & 943313817600\\
($4^2$,1) & 983657364480\\
(4,3,2) & 2002024200960\\
(4,3,$1^2$) & 2379988396800\\
($3^3$) & 765174574080\\
\hline
\end{tabular}
\end{center}
\end{table}


\section*{Acknowlegement}
The work of HdG is supported by NSERC of Canada.
Part of this work was completed by JD under the NSERC USRA program.


\section*{References}

\bibliography{wein_refs}
\bibliographystyle{iopart-num}

\end{document}